\documentclass[12pt,a4paper]{article}
\usepackage{graphicx}
\begin{document}
\def\curl{\mathop{\rm curl}\nolimits}
\def\dvg{\mathop{\rm div}\nolimits}
\newcommand{\abl}[2]{\frac{\partial #1}{\partial #2}}
\newcommand{\subabl}[1]{\frac{D #1}{D\,t}}
\newfont{\mmitb}{cmmib10}
\title{A well posed acoustic analogy based on a moving acoustic medium\thanks{Presented at the Aeroacoustic workshop SWING in Dresden 1999. Modified only by adding the abstract and this footnote.}}
\author{ Willi M\"ohring, Max-Planck-Institut f\"ur Str\"omungsforschung, \\
Postfach 2853, D-37018 G\"ottingen}
\date{}
\maketitle
\begin{abstract}
 For flows of a lossless gas, the stagnation enthalpy obeys a linear convected wave equation with coefficients which depend on the flow variables. This equation is self-adjoint and one has a reciprocity relation between source and observer. It fulfills for subsonic flow a quadratic conservation equation implying stability. It is taken as basis for an acoustic analogy and is applied to the sound generation by the collision of a convected vortex and a rigid cylinder.
\end{abstract}

\section{Introduction}

The concept of an acoustic analogy was introduced by Lighthill \cite{1}.
This idea offered him the possibility to derive important results on the
generation of aerodynamic noise without relying on expansions or perturbation
schemes. This is especially important, as the acoustic energy generated by an
unsteady flow field represents usually only a minute part of the energy flux
occurring in the flow and errors which are small compared to the flow
quantities may be very large if one compares them with the sound quantities.
The conditions under which perturbation schemes based on small sound
level or on small flow Mach number may be used, have been clarified in later
researches \cite{1a,2a,3a}, although there still remain open questions.

Lighthill derived without any approximations an exactly valid equation, which
admits an acoustic interpretation. He showed that every compressible flow
fulfills an inhomogeneous wave equation with a quadrupole type source
distribution. As the inhomogeneous wave equation describes the generation and
propagation of sound in an ideal acoustic medium at rest, he had related
aerodynamic noise to sound waves in the ideal acoustic medium. This relation
is exactly true, but it relates the effects of convection and refraction by a
steady basic flow to a source distribution in an acoustic medium without flow
and this is difficult to visualize. One is therefore inclined to look for a
relation to an ideal acoustic medium in motion.

Here we will identify the operator which describes the propagation of sound in
irrotational, isentropic flow and we will base the analogy on that operator.
This of course does not mean, that vortices or entropy inhomogeneities are
excluded. It is very similar to the situation found in Lighthill's analogy.
There, flows were exluded from the medium and they occurred as sources. Here,
vortices and entropy inhomogeneities are excluded and they occur as sources.
The equation is therefore very well suited to study the sound radiation from
vortices which are convected in an irrotational flow. The acoustic variable
which we use is the stagnation enthalpy. This variable has been used with
aeroacoustic applications in mind before \cite{5a,5b}. The equations
considered before differ however from that derived here, but compare
\cite{6a}. The operator which replaces the wave operator here is a
self-adjoint one, even if the flow field is completely arbitrary. It is not
necessary that it is irrotational or that it fulfills the basic eqations of
fluid dynamics. The self-adjointness leads to a reciprocity principle which is
valid for an exchange of source and observer. It also implies the existence of
a variational principle from which the basic equation of the analogy can be
derived. This principle is the simplest generalization of the variational
principle for the wave equation, partial derivatives with respect to time are
replaced by material derivatives, which seems natural if one requires
Galilei-invariance. From the variational principle one can then conclude the
validity of an energy theorem. The main drawback is of course the lack of a
simple explicit expression for the Green's function. Notice however, that for
certain important situations a Green's function to first order has been
determined in \cite{10a}. An alternative would of course be a numerical
solution. To check its feasibility, a general purpose PDE-solver for a PC has
been applied to calculate the sound generated by a two-dimensional vortex
convected along a circular cylinder.

\section{The acoustic analogy}
\subsection{Preliminaries}
Lighthill based his theory of aerodynamic sound generation on an eqaution,
which he obtained by cross-differentiation from the Euler equations and from
the continuity equation
\begin{eqnarray*}
\abl{{\bf \rho v}}t + \nabla\cdot \rho {\bf v\, v} + \nabla p & = & 0 \\
\abl{\rho}t + \nabla \cdot \rho {\bf v} & = 0,
\end{eqnarray*}
namely
\begin{equation}
\Delta p -\frac1{a^2} \frac{\partial^2 p}{\partial t^2} +
\nabla\cdot\nabla\cdot \rho  {\bf v\; v} = 0.  \label{gl1}
\end{equation}
Here $\rho$ denotes the density, $p$ the pressure, $a$ the speed of sound, and
${\bf v}$ the particle velocity. This equation is valid for isentropic flow if
losses and temporal variations of the speed of sound can be ignored. Temporal
averages can be subtracted, therefore one may assume that only fluctuating
quantities are contained in eq.(\ref{gl1}). No linearization has been
performed in the derivation. Therefore eq.(\ref{gl1}) is valid under very
general conditions. Often, especially for low Mach number flows, one neglects
the acoustic contributions in the double divergence in eq.(\ref{gl1}) and
considers this term as a known source term for the sound generation. Then one
describes the sound field as a wave obeying the wave equation, i.e. as a sound
field propagating in a non-moving medium. Convection and refraction effects
are then neglected. Eq.(\ref{gl1}) is however generally true and these effects
are in principle contained in eq.(\ref{gl1}). Attempts to extract them have
been made by separating the velocity into a mean and fluctuating part and
thereby  to obtain these effects. Although it should be possible to describe
mean flow effects in this way, it has been felt that a more apropriate
description should be obtained through a modification of eq.(\ref{gl1}).
Instead of the wave operator, a "convected wave operator" which describes the
propagation of sound in a moving medium seems more apropriate. One then
rewrites all fluid quantities as a superposition
\[ {\bf v} \rightarrow {\bf v}_0 + {\bf v}, p \rightarrow p_0 + p, \dots \]
of "nonacoustic" and "acoustic" variables and rewrites the basic equations
in terms of these variables. If one neglects contributions which are nonlinear
in the non-subscribed variables, one obtains the acoustic equations. It is
well known, that one can derive from this system one "convected wave equation"
for unidirectional (in $x$-direction) shear flow or for potential flows,
namely
\[
{\cal L}_{\rm shear}\, p = \frac{1}{a_0^2} \subabl{} \nabla \cdot a_0^2\nabla p
- \frac{1}{a_0^2}\frac{D^3 p}{D\,t^3} -
2 (\nabla u_0) \cdot\nabla \,\abl{p}{x}  =0 \quad {\rm with} \quad
\frac{D}{D\, t}=\frac{\partial}{\partial t} + {\bf v} \cdot \nabla
\]
for unidirectional and with an acoustic potential $\phi$
\[
{\cal L}_{\rm pot}\, \phi =
\nabla\cdot (\rho_0 \nabla \phi - \frac{\rho_0}{a_0^2}\subabl{\phi}{\bf v}_0)
- \abl{\,}{t}\frac{\rho_0}{a_0^2} \subabl{\phi} = 0\]
for irrotational flows. For aerodynamic noise one should not neglect the
nonlinear terms. It is however possible, very similar to Lighthill's approach,
to collect the linear and nonlinear terms and to derive an inhomogeneous wave
equation. It is even possible -- and it has been done e.~g. by Tam and Auriault
\cite{LiTa} -- to follow this approach in the full continuity and Euler
equations and derive an inhomogeneous linear system of equations. Then one
could use for ${\bf v}_0$, $p_0$, etc. the temporal mean values. The equations
which are obtained by this method are much more complicated than the wave
equation and are usually solvable only numerically. Furthermore they are
rather different from the equations usually studied in mathematical physics
and little is known about existence and uniqueness of solutions. It is however
known, that the linear parts of the equations agree with the stability
equations and as many flows are unstable, one has to expect, that the
equations will be unstable. Of course, exponentially growing solutions are
physically excluded -- at least for longer times -- and therefore they do not
occur in the correct solutions. Small errors will however produce these
instabilities and provisions have to be made to limit their growth. How these
provisions influence the sound obtained from these calculations is difficult
to assess.

Here we follow a different strategy which is related to the propagation of
sound waves in potential flows, but differs from the above described method
significantly. We do not separate the velocity into an irrotational part and a
remainder but follow a path used by Howe \cite{5a}. He observed that
Bernoulli's equation states that the stagnation enthalpy differs from the
potential only by a sign and by a time derivative. For irrotational sound
waves one could therefore use the stagnation enthalpy instead of the
potential. The stagnation enthalpy is however defined also for rotational
flows and could be used as a variable of an acoustic analogy for arbitrary
flows. This is what we will do. We will derive a convected wave equation for
the stagnation enthalpy. We will discuss its main properties and show that it
possesses many of the formal properties of the ordinary wave operator. An
important example is an energy conservation theorem with an energy density
which is positive for subsonic flows. This excludes instabilities, the
influence of small errors remains small. The provisions which are necessary in
many other analogies to limit the troublesome growth of instabilities are not
necessary here.

\subsection{Basic Relations}
To obtain the equations
for the acoustic analogy including convection effects, one starts from the
Euler equation for compressible flow. Crocco's form of these equations reads
\begin{equation} \frac{\partial {\bf v}}{\partial t} + \nabla B = -{\bf L},
\qquad {\bf L} = {\mmitb \omega} \times {\bf v} - T \nabla s, \quad
{\mmitb \omega} = \curl {\bf v}.\label{a1}
\end{equation}
$B$ denotes the total
enthalpy $B=h+\frac 12 {\bf v}^2$ with the enthalpy $h$ and the velocity
${\bf v}$. $T$ is the temperature und $s$ the entropy. From the energy theorem
one finds for the total enthalpy
\begin{equation} \frac{D\, B}{D\, t} = \frac1{\rho}
\frac{\partial p}{\partial t}\label{a2}
\end{equation}
If one writes $ d \rho = a^{-2} d\,p + \rho_sd\,s $ where  $\rho_s$ denotes
the derivative of the density with respect to the entropy, one gets from the
continuity equation
\begin{equation} \frac{\rho}{a^2}\Bigl(\frac{D\, B}{D\, t}\Bigr) +
\dvg {\bf w} = - \rho_s \frac{\partial s}{\partial t}=q_s,  \label{a3}
\end{equation}
where ${\bf w}$ denotes the mass flux. If one multiplies
Crocco's vortex theorem (\ref{a1}) with the density $\rho$ one obtains for the
mass flux ${\bf w}$ the equation
\begin{equation}
\frac{\partial {\bf w}}{\partial t} -
\frac{\rho {\bf v}}{a^2} \frac{D\, B}{D\, t} + \rho \nabla B =
-\rho {\bf L} + q_s {\bf v} = {\bf K}.        \label{a4}
\end{equation}
It is easy to eliminate the mass flux ${\bf w}$ from these equations. One then
obtains an equation, which is linear in $B$
\begin{equation} {\cal L}\,B =
\nabla \cdot (\rho \nabla B - \frac{\rho {\bf v}}{a^2} \frac{D\, B}{D\, t}) -
\frac{\partial}{\partial t} \frac{\rho}{a^2}\frac{D\, B}{D\, t} = -\dvg \rho
{\bf L}+\frac {\partial q_s}{\partial t} + \dvg q_s{\bf v} = q_{tot}.
\label{a5}
\end{equation}
The operator ${\cal L}$ obtained here agrees completetely with the operator
${\cal L}_{\rm pot}$ given in the previous section for the propagation of
irrotational sound waves. If one inserts the sources from the equations
(\ref{a3}) und (\ref{a4}), one finds
\begin{equation} q_{tot} =
\left(\frac{\partial}{\partial t}\rho_s \frac{\partial}{\partial t}+ \dvg
\rho_s {\bf v} \frac{\partial}{\partial t} + \dvg \rho T \nabla \right)s +
\dvg \rho {\bf v}\times \omega.         \label{aq}
\end{equation}
The sources are linear expressions in the vorticity vector and the entropy. In
this analogy one may think of the sound as being generated from vorticity and
entropy inhomogeneities.

With acoustical applications in mind, the total enthalpy was first used by Howe
\cite{5a} and recently proposed also by Doak \cite{5b}. A comparison with
their equations shows, that the equation for $B$ is not uniquely determined.
Howe's convected wave operator is
\[ {\cal L}_{\rm Howe}\, B = \Delta B -\frac1{a^2}\subabl{{\bf v}}\cdot\nabla B
-\subabl{\;}\frac1{a^2}\subabl{B} \]
and Doak's
\[{\cal L}_{\rm Doak}\, B = \Delta B -\frac1{a^2}
\Bigg[\frac{\partial^2 B}{\partial t^2} + \Bigg(2{\bf v}\abl{\;}{t} +
{\mmitb \omega} \times {\bf v} +T\nabla s - 2 \nabla h\Bigg)\nabla B +
{\bf v}{\bf v}\cdot\nabla\cdot\nabla B\Bigg].  \]
The right hand sides of these equations differ from the right hand side
of eq.(\ref{a5}). We will restrict ourselves to eq. (\ref{a5}). Notice however
that the principle parts -- i.e. those terms which contain second derivatives
of $B$ -- of the three convected wave operators agree. This means that they
agree in the high frequency limit of geometric acoustics and agree with the
well known geometric acoustic theory.

A certain simplification of eq. (\ref{a5})  is possible if one requires
that $\rho$ and ${\bf v}$ fulfill the continuity equation. One then gets
\begin{equation}
{\cal L}\,B =\nabla \cdot (\rho \nabla B )
- \rho \frac{D}{D t} \frac{1}{a^2}\frac{D\, B}{D\, t}.
\label{a5a}\end{equation}

The equation (\ref{a5}) or (\ref{a5a}) is a generalization of the wave
equation. It reduces to the wave equation if one assumes in (\ref{a5}) ${\bf
v} =0$  and if one assumes further, that $\rho$ and $a^2$ are constant. The
equation agrees then with Lighthill's equation in the form of Powell
\cite{4a}. It seems to be a rather complicated equation. Considering a
variational principle we will however see, that equation (\ref{a5},\ref{a5a})
is actually the simplest equation which contains a flow velocity.
A simplification of the source is possible for an ideal gas. For an
ideal gas  $\rho$ is the product of two functions which depend only on $p$
and on $s$. Then one has $\rho_s = \rho f(s)$ with some function $f(s)$ and
one obtains for the sources of eq. (\ref{a5})
\[q_{tot}=-\dvg \rho {\bf L}+\rho \frac{D}{Dt}f(s)\frac{\partial s}{\partial t}
= - \dvg \rho {\bf L}-
\rho \frac{\partial {\bf v}}{\partial t}\cdot f(s) \nabla s
\]
if one makes use of the entropy conservation law.

The equation (\ref{a3}) and (\ref{a4}) rsp. (\ref{a5}) are now considered as
the basic equations of the acoustic analogy. They agree formally with the
linearized equations which describe small perturbations of a steady potential
flow. The zeroth order equations are then given by $B_0=0$ (because of the
Bernoulli equation) and $\dvg {\bf w}_0 =0$, i.e. the zeroth order versions of
the eqs. (\ref{a3}) and (\ref{a4}). The first order eqs. of (\ref{a3}) and
(\ref{a4}) are then obtained, if the fields $\rho$, $a$, and ${\bf v}$ are
replaced by their zero order values. Notice that $\rho {\bf v}$ in eq.
(\ref{a4}) then becomes $\rho_0 {\bf v}_0$ and it differs from ${\bf w}$ which
becomes ${\bf w}_1$. In that sense the left hand side of the eqs. (\ref{a3})
and (\ref{a4}) rsp. (\ref{a5}) are considered as equations which describe
sound propagation in potential flows. This is very similar to Lighthill's
interpretation of the wave equation as an equation which describes sound
propagation in a medium at rest. This interpretation is valid only if $\rho$,
$a$, and ${\bf v}$ are steady fields which fulfill the equations for
compressible irrotational flow. We will however not require this, as ${\cal
L}$ is also well defined without this assumption and there are
no advantages in assuming it. The eqs. (\ref{a3}) and (\ref{a4}) rsp.
(\ref{a5}) are exactly valid identities. They are considered in the following
as a system of linear partial differential equations for the variables $B$ and
${\bf w}$. This implies also that the right hand sides of the equations
(\ref{a3}) and (\ref{a4}) rsp. (\ref{a5}) are considered as the sources of the
sound. They are related to vortices and entropy inhomogeneities. This seems
reasonable if one wants to study the generation and propagation of sound in a
potential flow. Then the flow consists of a superposition of an irrotational
steady part and an unsteady part. Often one will superpose these two
contributions linearly. We will not require that. For the steady flow one has
a constant value of the total enthalpy  $B$. One may then assume $B$ to be
zero. $B$ is then solely related to the unsteady part and is small if this
part is small. In irrotational flow there is a potential $\Phi$. This obeys
the Bernoulli equation
\begin{equation}
\Phi_t + B = 0 \label{a6}
\end{equation}
$B$  then differs in regions where the flow is irrotational from the temporal
derivative of the potential only by its sign, it is however -- contrary to the
acoustic potential defined everywhere.

This equation (\ref{a5}) was originally derived in \cite{6a}.  Let us now
derive its main properties. The first important point is
that ${\cal L}$ is formally self-adjoint. This follows from the
fact that one has for arbitrary functions $B$ and $\tilde B$ the identity
\begin{equation}
\tilde B {\cal L} B - B {\cal L} \tilde B =
\frac{\partial l_0}{\partial t} + \frac{\partial l_i}{\partial x^i}
\label{b0} \end{equation}
with
\begin{equation}
l_0=-\frac{\rho}{a^2}\Bigl(\tilde B \frac{D\, B}{D\, t} -
B \frac{D\, \tilde B}{D\, t} \Bigr),\quad
l_i=\rho \Bigl(\tilde B \frac{\partial B}{\partial x^i} -
B \frac{\partial \tilde B}{\partial x^i} \Bigr)+l_0 v_i . \label{b01}
\end{equation}
This equation is easy to check. It implies also that one has for a scalar
product $(f,g)$ defined by $(f,g)=\int f\, g \, d^3xdt$ the relation
\begin{equation}
(\tilde B,{\cal L}B) = (B,{\cal L}\tilde B) \label{a7}
\end{equation}
if $B$ and $\tilde B$ vanish on the boundary of the integration region or
decay sufficiently rapidly at infinity.
\subsection{Reciprocity}
One may derive from the symmetry in eq. (\ref{a7}) a reciprocity relation. To
be specific let $G({\bf x},t,{\bf y},t')$ be the Green's function associated
with ${\cal L}$, i.e.
\begin{equation}
{\cal L} G({\bf x},t,{\bf y},t') =-\delta({\bf x}-{\bf y})\, \delta(t-t'),
\quad {\rm with} \quad G({\bf x},t,{\bf y},t') =0 {\quad} {\rm for} \quad
t<t', \label{b1}
\end{equation}
where we have assumed that $G$ is causal, i.e. it vanishes for all times
before the source is switched on, which occurs at $t=t'$. There exists also an
advanced Green's function $G_{\rm adv}$ with
\begin{equation}
{\cal L} G_{\rm adv}({\bf x},t,{\bf y},t') =-\delta({\bf x}-{\bf y})\,
\delta(t-t'), \quad {\rm with} \quad G_{\rm adv}({\bf x},t,{\bf y},t') =0
{\quad} {\rm for} \quad
t>t'. \label{b2}
\end{equation}
It is easy to see that eq. (\ref{a5}) is invariant with respect to time
reversal $t \rightarrow -t$ if at the same time the sign of the velocity is
reversed. Therefore time reversal transforms a Green's function into a Green's
function. As the time reversal interchanges the inequalities $t>t'$and $t>t'$
one has
\begin{equation}
G({\bf x},-t,{\bf y},-t';-{\bf v}({\bf x},-t)) =
G_{\rm adv}({\bf x},t,{\bf y},t';{\bf v}({\bf x},t)) \label{b3}
\end{equation}
where we have added for clarity the function ${\bf v}({\bf x},t)$ to the list
of arguments of the Green's function. In addition the functions $\rho({\bf
x},t)$ and $a({\bf x},t)$ have to be replaced by $\rho({\bf x},-t)$
and $a({\bf x},-t)$.

One may now apply eq.(\ref{a7}) with
\[ B =G({\bf x},t,{\bf y},t') \quad {\rm and} \quad
 \tilde B =G_{\rm adv}({\bf x},t,{\bf z},t'') \]
and one obtains
\begin{equation}
( G_{\rm adv}({\bf x},t,{\bf z},t''), {\cal L}  G({\bf x},t,{\bf y},t'))
=( {\cal L} G_{\rm adv}({\bf x},t,{\bf z},t''),G({\bf x},t,{\bf y},t'))
\label{b4}\end{equation}
Let us indicate briefly that there are no contributions from the right hand
side of eq. (\ref{b0}). If the integration in eq. (\ref{b4}) is performed over
a large cylinder in the ${\bf x}$,$t$-space which extends over a large sphere
in ${\bf x}$-space and over all $t$ with $T_0<t<T_1$, one has surface
contributions which are to be evaluated over the large sphere at $t=T_0$ and
at $t=T_1$ and over the surface of the large sphere in ${\bf x}$-space for all
$t$ with $T_0<t<T_1$. If $T_0$ is before $t'$ and $t''$ and $T_1$ after $t'$
and $t''$ there are no contributions from the space integrals at $t=T_0$ and
at $t=T_1$, as at least one factor vanishes in $l_0$ and in the $l_i$, namely
the factors containing $G_{\rm adv}$ at $t=T_1$ and those containing $G$ at
$t=T_0$. There is also no contribution from the large surface in ${\bf
x}$-space if it is selected so large that no signal which was generated at
$t=t'$ and at ${\bf x}={\bf y}$ has reached this surface. Therefore eq.
(\ref{b4}) is true, and one can evaluate the scalar products with the
$\delta$-functions in the eqs. (\ref{b1}) and (\ref{b2}) easily, and one
obtains the equation
\begin{equation}
 -G_{\rm adv}({\bf y},t',{\bf z},t'') =-G({\bf z},t'',{\bf y},t')
\label{b5}\end{equation}
which can with eq. (\ref{b3}) be rewritten as
\[
G({\bf y},-t',{\bf z},-t'';-{\bf v}({\bf y},-t)) =
G({\bf z},t'',{\bf y},t';{\bf v}({\bf z},t)).\]
This is the reciprocity principle with reversed flow.
\subsection{Variational Principle and Energy Conservation}
From the self-adjointness
on can conclude the existence of a variational principle from which  eq.
(\ref{a5}) can be derived. One has

\begin{equation}
\delta L = 0 \quad {\rm mit} \quad L=\frac 12 (B,{\cal L} B) -(B,q_{tot})
\label{a8} \end{equation}
as
\begin{equation}
\delta L =\frac 12 (\delta B,{\cal L} B) + \frac 12 (B,{\cal L} \delta B )
-(\delta B,q_{tot})= (\delta B,{\cal L} B -q_{tot}).
\label{a9} \end{equation}

The Lagrangian from (\ref{a8}) can be simplified somewhat, if the second
derivatives in ${\cal L}$ are eliminated with integration by parts. One finds
then
\begin{equation}
L=\int\int \Bigl[\frac{\rho}{2a ^2} \bigl(\frac{D\, B}{D\, t}\bigr)^2
-\frac {\rho}2(\nabla B)^2- q_{tot} B \Bigr]
d^3x\,dt.
\label{a10a}\end{equation}

It seems very remarkable that this variational principle seems to be the
simplest possible extension of the well known principle for the wave equation
which is invariant with respect to Galilei-transformations. The Lagrangian in
eq. (\ref{a8}) differs from the Lagrangian of the wave equation only by the
fact, that partial derivatives with respect to time are replaced by material
derivatives formed with the velocity field ${\bf v}$.

Now it is possible to obtain from a variational principle an energy theorem,
rsp. an energy conservation law, if the Lagrangian density does not depend
explicitly from the time. If $l$ denotes the  Lagrangian density from eq.
 (\ref{a10a}), i.e.
\begin{equation}
l=\frac{\rho}{2a ^2} \bigl(\frac{D\, B}{D\, t}\bigr)^2
-\frac {\rho}2(\nabla B)^2 - q_{tot} B,
\label{a10}\end{equation}
one has for the energy theorem
\begin{equation}
\frac{\partial}{\partial t}\Bigl(\dot B\frac{\partial l}{\partial \dot B} -l
\Bigr)
+\frac{\partial}{\partial x^i} \dot B\frac{\partial l}{\partial B_{x^i}}
= \frac{\partial l}{\partial t}
\label{a11} \end{equation}
where the time derivative on the right hand side acts only on the explicit
time dependance in $l$, i.e. here in $\rho$, $a$, ${\bf v}$ and $q_{tot}$.
The variable $B$ and its derivatives are to be kept constant. The time
derivative on the left hand side of eq.(\ref{a11}) acts also on the implicit
dependance in $B$ and its derivatives $\dot B$ und $B_{x^i}$. Only the spatial
coordinates $x^i$ are to be kept constant. An energy conservation law is
obtained from (\ref{a11}) if the Lagrangian density does not contain the time
explicitly. In general one obtains for the energy flux
${\displaystyle U_i= \dot B\frac{\partial l}{\partial B_{x^i}}}$
and the energy density
${\displaystyle e= \dot B\frac{\partial l}{\partial \dot B} -l}$
the explicit expressions
\begin{equation}
{\bf U} = \rho \dot B \Bigl(\frac1{a ^2} \frac{D\, B}{D\, t} {\bf v}
- \nabla B \Bigr)
\label{a12} \end{equation}
and
\[
e = \frac{\rho}{a^2}\dot B \frac{D\, B}{D\, t}
-\frac{\rho}{2a ^2} \bigl(\frac{D\, B}{D\, t}\bigr)^2
+\frac {\rho}2(\nabla B)^2 =  \frac{\rho}{a^2} {\dot B}^2
+\frac {\rho}2(\nabla B)^2 -
\frac{\rho}{2a^2}\Bigl({\bf v} \cdot \nabla B \Bigr)^2
\]
which shows that the energy density $e$ is positive for subsonic ${\bf v}$.
The energy theorem is then of the form
\begin{displaymath}
{\partial e \over \partial t} + \dvg {\bf U} =q_{\rm En}
\end{displaymath}
with an energy density $e$ and an energy source densitye $q_{\rm En}$.
A useful relation is obtained if one integrates this equation over the time
for finite time events or if one averages this equation for the case of steady
phenomena. If one denotes the resulting quantities by an overbar, one obtains
\begin{equation}
\dvg {\bar{\bf U}} ={\bar q_{\rm En}}.
\end{equation}

Another important conclusion can be drawn from the energy theorem if one
applies it to the solution of an initial value problem with vanishing right
hand side $q_{tot}$. If one assumes that the solution vanishes for large
$|{\bf x}|$, one obtains
\[ \abl{\;}{t}\int e\, d^3\, x \; = 0,\]
i.e. the toatal energy in the sound field remains constant. As it is a sum of
positive contributions, none of these -- e.g. $\dot{B}$ -- can grow
exponentially in time, i.e. instabilities cannot occur.

The physical meaning of this energy flux becomes clearer if one considers an
irrotational isentropic region. There on may write (\ref{a12}) with the eqs.
(\ref{a2}) and (\ref{a6}) as
\begin{equation}
{\bf U} =  \dot B \Bigl(\frac 1{a ^2} \dot p {\bf v}+ \rho \nabla \Phi_t \Bigr).
\label{a13}\end{equation}
One may compare this energy flux with the flux from the  Blokhintzev energy
theorem which is valid for the propagation of irrotational sound waves in an
irrotational mean flow in linear approximation. One has neglected quantities
which are quadratically in the acoustical quantities. Notice that no
linearization has been assumed in the derivation of the energy theorem
(\ref{a10},\ref{a11}). It is insofar an exact identity, only dissipative
effects have been ignored. If one denotes in the Blokhintzev energy flux with
${\bf U}_{\rm Bl}$, the density, the speed of sound and the velocity of the
irrotational mean flow with $\rho_0$, $a_0$ and ${\bf v_0}$ and with $p'$ and
${\bf \phi'}$ the acoustic pressure and the acoustic potential, one can write
\begin{equation}
{\bf U}_{\rm Bl} = (\frac 1{\rho_0} p' + {\bf v_0} \cdot \nabla \phi')
(\frac 1{a_0^2} p'{\bf v_0} + \rho_0\nabla \phi').
\label{a13a}\end{equation}
A comparison of eq. (\ref{a13a}) with eq. (\ref{a13}) shows, that both energy
fluxes are products of two factors, where the factors of eq. (\ref{a13}) are
just the time derivatives of the factors of the Blokhintzev energy flux
(\ref{a13a}). If one thinks of the sound field as a superposition of temporal
Fourier modes, one finds that the average energy flux consists of a
superposition from fluxes of the modes. In the energy flux of eq. (\ref{a12})
all contributions contain an additional factor $\omega^2$ if $\omega$ denotes
the angular frequency of the Fourier modes.
\subsection{Solutions}
As a first application one may consider the case of constant values of the
velocity ${\bf v}$, density $\rho$ and speed of sound $a$. Then vorticity and
entropy inhomogeneities are convected with the velocity ${\bf v}$. These
inhomogeneities and also the total enthalpy $B$ are then functions of
${\bf x} - {\bf v} t$ only, i.e. $B= B(${\bf x} - {\bf v} t$)$,
$s= F_s(${\bf x} - {\bf v} t$)$
and $\omega = {\bf F}_{\omega}(${\bf x} - {\bf v} t$)$.
Then eq. (\ref{a5}) leads to
\begin{displaymath}
\rho \Delta B = q_{tot}, \qquad
q_{tot} =
\left(\frac{\partial}{\partial t}(\rho_s \frac{\partial}{\partial t}+
 \dvg \rho_s {\bf v} \frac{\partial}{\partial t})
 + \dvg \rho T \nabla \right)F_s + \dvg \rho {\bf v}\times F_{\omega}.
\end{displaymath}
One may then introduce $B_{\omega}$ and $B_s$ by
\begin{displaymath}
\rho \Delta B_{\omega} = F_{\omega} \quad {\rm und} \quad
\rho \Delta B_s = F_s
\end{displaymath}
and one obtains
\begin{equation}
B=
\left(\frac{\partial}{\partial t}(\rho_s \frac{\partial}{\partial t}+
 \dvg \rho_s {\bf v} \frac{\partial}{\partial t})
 + \dvg \rho T \nabla \right)B_s + \dvg \rho {\bf v}\times B_{\omega}.
\label{a13b}\end{equation}
This shows -- as one might have expected -- that passively convected entropy
and vorticity inhomogeneities do not radiate sound. In the general case these
quantities will not be passively convected and one needs for its determination
extra equations. For the entropy one may use the equation of entropy
conservation
\begin{displaymath}
\frac{D\,s}{D\, t} =0
\end{displaymath}
and for the vorticity the Beltrami vortex theorem
\begin{displaymath}
\frac{D}{D\, t} {\omega \over \rho} = {\omega \over \rho} \cdot \nabla
{\bf v}.
\end{displaymath}
Here we consider especially the two-dimensional case. Then the right hand side
of the Beltrami vortex theorem, which is related to the stretching of vortex
lines, vanishes.

In addition to the differential equations one needs boundary conditions. If
one is interested in cases where the vorticity vanishes at solid walls in the
flow region, one may use the relation (\ref{a6}) between total enthalpy $B$
and potential $\Phi$ and one finds that the normal component of the velocity
vanishes at a fixed surface if the normal derivative of $B$ vanishes there.
With eq. (\ref{a12}) one notices that the normal component of the energy flux
(\ref{a12}) vanishes at rigid walls if the normal component of the velocity
${\bf v}$ vanishes there. One would expect this of course.

In addition one needs conditions of no-reflexion at the boundary of the
computation region. We will here assume the simplest quasi-onedimensional
condition and require there
\begin{displaymath}
\frac{\partial B}{\partial t} = -({\bf v} + {\bf n} a)\cdot \nabla B,
\end{displaymath}
where ${\bf n}$ denotes the outer normal of the computation region.

As a numerical example, we consider a localized vortex of radius 1 and of
vanishing total vorticity which is convected in a flow around a circular
cylinder of radius $1/2$ and situated at $x=0$ and $y=0$. The initial azimuthal
velocity $w$ around the center of the vortex, which is situated initially at
$x_0=-3$, $y_0=0.5$, is assumed to be
\[ w=(1-4r^2)(1-r^2)^2, \]
$r$ denotes the distance from the vortex center. For the velocity, we assume
an incompressible potential flow of velocity 1 in $x$-direction at $x=-\infty$.
The density is chosen as 1, the speed of sound as 2. Initial values for $B$
are obtained from eq.(\ref{a13b}) and are given by
\[B=0.5 (y-y_0)(1-r^2)^3 \quad {\rm for} \quad r<1. \]
This problem is treated with the general purpose PDE-solver PDEase/2 which
runs on a PC. A grayplot of the $B$-field at two different times is shown in
figure 1.

\begin{figure}[ht]
\begin{center}
\includegraphics[width=5cm]{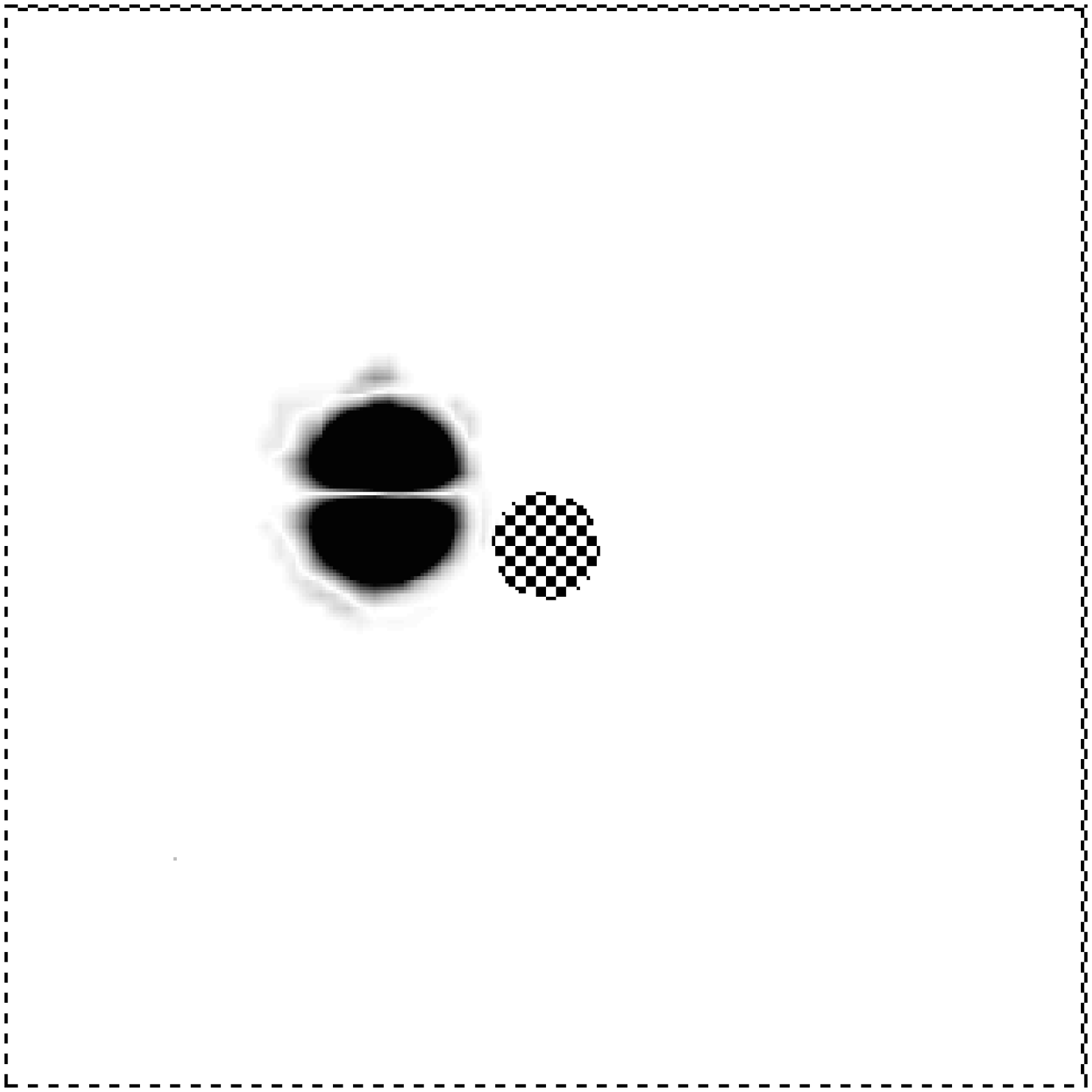}
\qquad
\includegraphics[width=5cm]{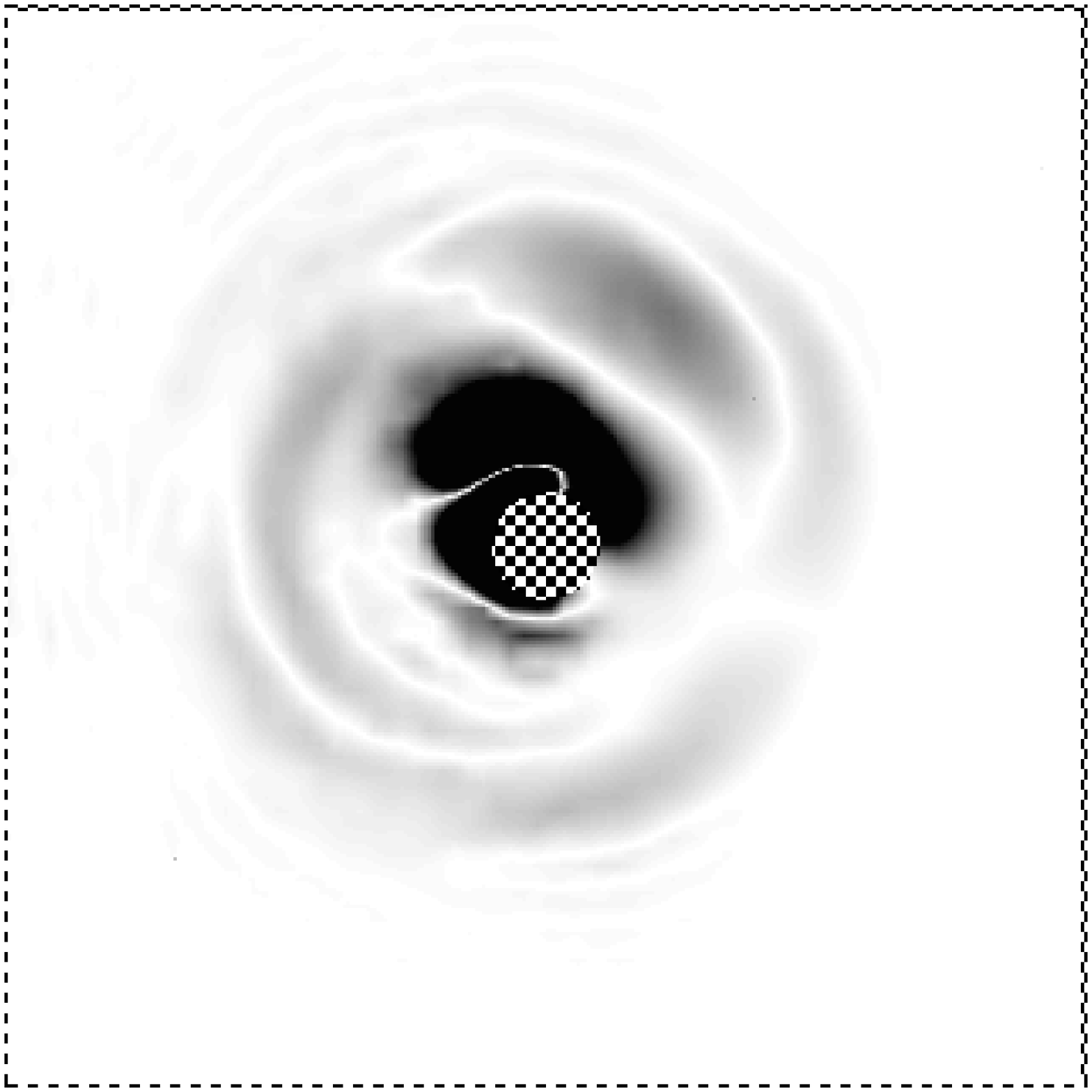}
\end{center}
\hskip1cm
Figure 1. The $B$-field generated by a vortex convected along a rigid
cylinder. Grayscales correspond to $|B|$, white to $B=0$.
\end{figure}
The left half shows a very early stage with the vortex to the left
of the cylinder. As the vortex is inserted in an inhomogeneous velocity field,
sound radiation begins immediately. The right half shows a later stage, where
the vortex has approached the cylinder. As the flow Mach number is not small,
one notices significant deviations from a dipole character. Hydrodynamic
instabilities are not observed, but numerical small-scale errors are
noticeable. A reliable numerical solution of eq.(\ref{a5}) requires obviously
more efforts.

\end{document}